# Achievement of 1 MA Discharges in Hydrogen-Boron Plasmas on EXL-50U

Yuejiang SHI(石跃江)[1,2,*], Xianming SONG(宋显明)[1,2], Dong GUO(郭栋)[1,2], Xinchen JIANG(姜欣辰)[1,2], Yong Liu(刘勇)[1,2], Xiang GU(顾翔)[1,2], Chengyue Liu (刘成岳)[3], Jia LI(李佳)[1,2], Shaodong SONG(宋绍栋)[1,2], Yumin WANG(王嵎民)[1,2], Songjian LI(李松建)[1,2], Renyi TAO(陶仁义)[1,2], Kun Han(韩坤)[1,2], Yong Lili DONG(董力立)[1,2], Cong ZHANG(张聪)[1,2], Yujin Li (李玉进)[1,2], Tiantian SUN(孙恬恬)[1,2], Fan GAO(高凡)[1,2], Yihang ZHAO(赵毅航)[1,2], Yupeng ZHANG(张宇鹏)[1,2], Quan Wu(武全)[1,2], Yan Yu(于岩)[1,2], Lei Li(李磊)[1,2], Guang YANG(杨光)[1,2], Xian Li(黄贤礼)[1,2], Hongda HE(何宏达)[4], Dong Li (李栋)[4], Huasheng XIE(谢华生)[1,2], Bing Liu (刘兵)[1,2], Hanyue Zhao (赵寒月)[1,2], Enwu YANG(杨恩武)[1,2], Yuanming YANG(杨圆明)[1,2], Yu WANG(王宇)[1,2], Lei HAN(韩磊)[1,2], Bo XIN(邢博)[1,2], Zhenxing WANG(王振兴)[1,2], Peihai ZHOU(周培海)[1,2], Wenwu LUO(罗文武)[1,2], Chao WU(吴超)[1,2], Xin ZHAO(赵鑫)[1,2], Chen Bian (边晨)[1,2], Kun Wu (吴堃)[1,2], Yunfeng LIANG(梁云峰)[1,5], Jiaqi DONG(董家齐)[1,2], Baoshan YUAN(袁保山)[1,2], Y-K Martin PENG(彭元凯)[1,2], Minsheng LIU(刘敏胜)[1,2], the EXL-50U team

1. Hebei Key Laboratory of Compact Fusion, Langfang 065001, People's Republic of China
2. ENN Science and Technology Development Co., Ltd., Langfang 065001, People's Republic of China
3. Hefei University of Technology, Hefei 230009, People's Republic of China
4. Southwestern Institute of Physics, Chengdu 610041, People's Republic of China
5. Forschungszentrum Jülich GmbH, Institute of Fusion Energy and Nuclear Waste Management-Plasmaphysik, 52425, 62Jülich, Germany

[*]E-mail of corresponding author: yjshi@ipp.ac.cn

**Abstract**

One mega ampere (MA) plasma discharges were achieved on the EXL-50U spherical torus (ST) in hydrogen-boron (p-B) plasmas with toroidal fields up to 1 T at major radii up to 0.6m. A key innovation in these experiments was the use of a boron-rich fueling strategy, incorporating a high-concentration boron-containing gas mixture and real-time boron powder injection during the discharges. The boron content in the fueling reached 10%, representing the first publicly reported MA-class hydrogen-boron plasma with such a high boron concentration. Non-inductive plasma startup was achieved using electron cyclotron resonance heating (ECRH), and a rapid current ramp-up to 1 MA was realized through the synergistic use of ECRH and the central solenoid (CS). With 800 kW of ECRH power, core electron temperatures of up to 3 keV were attained. These results demonstrate the feasibility of producing high-performance hydrogen-boron plasmas in a spherical torus configuration and offer important physics and engineering insights for future reactor-scale applications — particularly in the areas of low loop voltage current start-up and real-time boronization techniques.

The EXL-50U [1] is China's first large spherical torus (ST) device with a toroidal field reaching 1 T at ENN Science and Technology Development Co. Ltd.. For a private enterprise such as ENN, aiming at commercializing fusion, the fast iteration and upgrade of device are key measures to verify the commercial feasibility as early as possible. The physics and engineering design of EXL-50U was initiated in the middle of 2022, with the facility assembly completed at the end of 2023. Non-inductive current drive and high performance hydrogen-boron plasma physics are the key physics issues of EXL-50U's experiments which are directly related to EHL-2's physics [2,3] and future ST reactors [4].

High-current operation is crucial for enhancing overall parameters and confinement performance, whether a conventional tokamak or an ST. Major STs like MAST/MAST-U [5,6] and NSTX/NSTXU [7,8], which operate at MA-level currents, have achieved remarkably high-parameter experimental results. The start-up and ramp-up of plasma current have been commonly driven by a toroidal electric field induced by the CS coils in a conventional tokamak. However, this causes engineering difficulties for the ST due to the limited space available within a narrow center post. Although some current STs' start up and ramp up can depend on CS drive, such a current drive strategy is unlikely to work on the EHL-2 and future ST reactor. At the same time, a low loop voltage start-up is a requirement for future superconducting ST or tokamak reactors. Boronization for obtaining high-performance discharges in metal walls is also a key ITER issue [9].

Following the first plasma in January 2024, the goal of plasma current in the first EXL-50U experimental phase is 500 kA. In the experiments in 2024, the EXL-50U achieved a plasma current of 586 kA using ECRH (Electron Cyclotron Resonance Heating) for non-inductive current start-up and a current ramp-up with the synergetic effect of ECRH and CS. Stable and repeatable plasma current were also obtained under 500kA in limiter configuration and 400 kA in divertor configuration. More detail about the EXL-50U device can be seen in ref. [1].

Although glow discharge boronization is a regular wall cleaning method for EXL-50U, during the 2024 experiments, the primary working gas remained hydrogen, with occasional additions of diborane gas during the discharge flat-top phase. In the latest experiments conducted in April 2025, the filling gas for the entire discharge duration on EXL-50U was a diborane-hydrogen mixture. The composition of the gas mixture is 30% diborane and 70% hydrogen. At the same time, boron powder was injected during the ramp-up and flat-top phases. It is worth mentioning that the real-time boronization in the EXL-50U is the first experiment of its kind conducted on a full-metal-wall spherical torus device with tungsten limiter. Boron content in the working fuel is not less than 10%.

Switching to a diborane-hydrogen gas mixture resulted in a significant enhancement of the plasma current ramp-up rate. As shown in figure.1, under identical ECRH power, toroidal magnetic field, and CS flux consumption, within 70 ms after CS activation, the plasma current ramp-up rate of diborane discharge exhibited a 78% enhancement compared to hydrogen discharge, rising from 1.9 MA/s to 3.4 MA/s. In addition to the diborane discharge significantly enhancing the plasma current ramp-up rate during the startup phase to reduce CS flux consumption, the following two measures were also critical to achieving high-parameter MA-level discharges in EXL-50U.

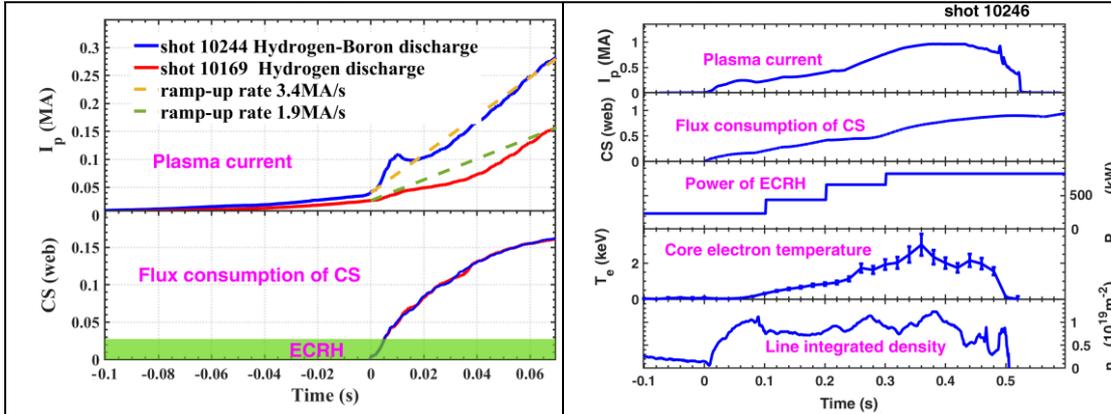

| Figure.1 Typical start-up waveforms in EXL-50U. 230kW 50GHz ECRH is applied from -0.2s for breakdown and startup. The toroidal field is 1 T at major radius = 0.6m. | Figure.2 The waveforms of 1MA discharge in EXL-50U. From top to bottom: plasma current, flux consumption of CS, power of ECRH, core electron density measured by Thomson scattering, line integrated density measured by HCN interferometers. A 50 GHz gyrotron operates continuously during the full discharge cycle, with three gyrotrons of 28 GHz operate in the stage from 0.1s. |

The first measure is that the four ECRH gyrotrons were switched on sequentially, with the ECRH power incrementally ramped up. This approach offers the advantage of applying lower ECRH power during the breakdown and start-up phases, which prevents excessive ECRH power from being unabsorbed, thereby minimizing impurity generation. As the plasma current and density increase, higher ECRH power levels are progressively applied to elevate plasma temperature and reduce CS flux consumption in ramp-up and flat-top phase.

The second is control of the plasma current ramp-up rate. It can be seen in figure.2 that current ramp-up rate undergoes a marked reduction after 70ms. The rationale for reducing the ramp-up rate is to prevent uncontrolled plasma expansion caused by rapid current growth, which would otherwise trigger intensive plasma-wall interactions (PWI) with low-field-side (LFS) limiters or first-wall components, ultimately leading to disruptive discharge termination.

By coordinating the above three measures, the EXL-50U device achieved stable and repeatable MA-level current discharge. The plasma current and position displacement strictly followed the predetermined waveform. Figure.3 shows the real plasma discharge image and reconstructed magnetic flux surface . The 1 MA plasma discharges on EXL-50U—characterized by an equilibrium-reconstructed peak plasma current of 1.03 MA—was achieved in a limiter configuration with an elongation of approximately 1.5 and an edge safety factor $q_a \sim 4$. The density and electron temperature profiles measured with Thomson scattering diagnostics are shown in figure.4. Under core density conditions around $1\times10^{19}m^{-3}$, the central electron temperature has reached 3 keV. In other MA-level, lower-density discharges, the highest electron temperature reached 3.5 keV.

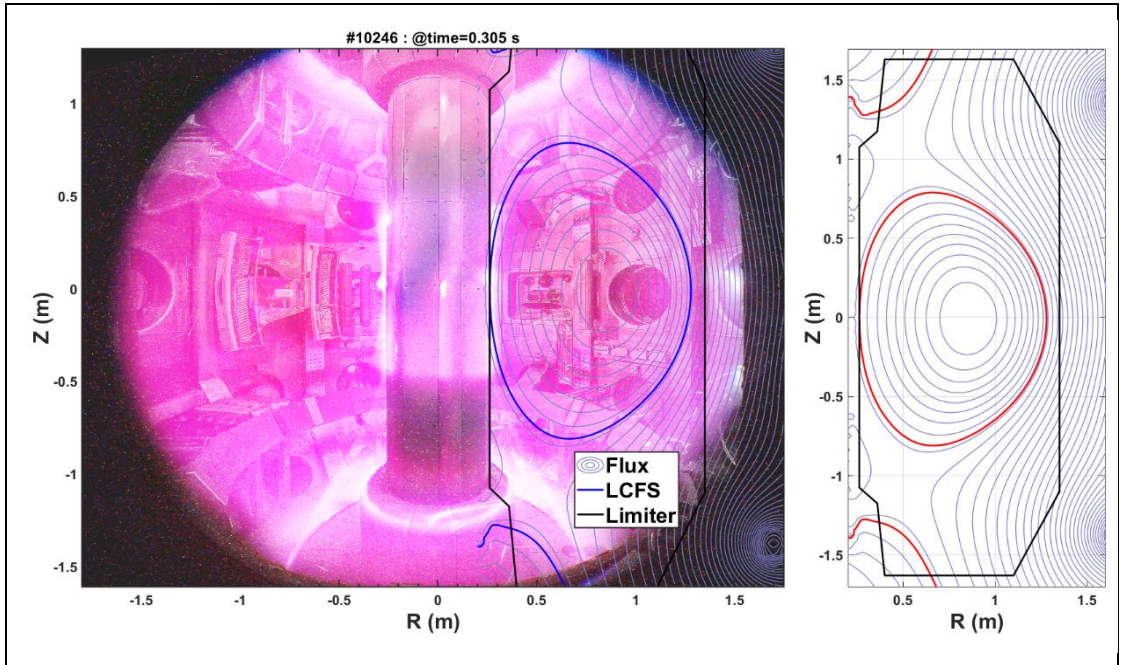

Figure.3 The plasma image and magnetic flux surface for 1MA discharge.

The relation between edge safety factor and plasma current $I_p$ is shown in figure.5. It is clear that higher currents are obtained under higher magnetic fields. On the other hand, the elongation of MA plasma in EXl-50U is a medium level. Therefore, under the current magnetic field level at 1T, the plasma current of EXL-50U still has room for further improvement with higher elongation discharges. On the other hand, EXL-50U plans to increase the magnetic field to 1.2T in 2026. The plasma current of EXL-50U will be further significantly increased in future experiment campaign.

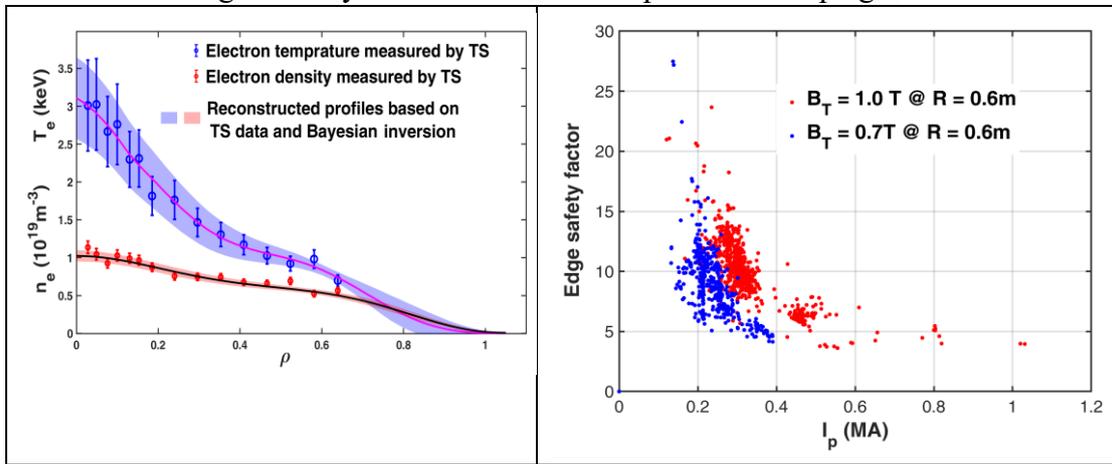

Figure.4 The electron temperature and    Figure.5 Relation between edge safety

| density profiles at t = 0.36s in shot 10246. | factor and plasma current. |

In summary, the MA-level discharges were achieved in hydrogen-boron plasmas for the first time on the EXL-50U, which demonstrated the scientific feasibility of generating high-performance hydrogen-boron plasmas in magnetic confinement device. The current drive mode of EXL-50U (ECRH non-inductive start-up, ECRH+CS synergy during the ramp-up phase) saves volt-seconds in the start-up and ramp-up phases, while maintaining a low loop voltage, which also provides the one possible solution for future large ST devices (such as EHL-2 and STEP) or superconducting tokamak reactor (such as ITER). The high parameters MA-level hydrogen-boron discharges on the EXL-50U shed light on the real-time boronization operation on future metal-wall magnetic confinement fusion reactor.

**Acknowledgments**
This work was supported by ENN Group and ENN Energy Research Institute. The authors would like to express their gratitude for the contributions of the ENN fusion center and collaborators in supporting the EXL-50U project.